\documentclass[iop,numberedappendix]{emulateapj}
\usepackage{epsfig}
\usepackage{xspace}
\usepackage{amsmath}
\usepackage{graphicx}

\renewcommand\micron{\mbox{$\mu$m}\xspace}
\newcommand{\msun}{$M_\sun$\xspace}

\newcommand{\secp}[1]{(\S\ref{#1})\xspace}
\newcommand{\sect}[1]{\S\ref{#1}\xspace}
\newcommand{\eqt}[1]{Eq.\ (\ref{#1})\xspace}
\newcommand{\eqp}[1]{(Eq.\ [\ref{#1}])\xspace}
\newcommand{\panp}[1]{({\em #1})\xspace}
\newcommand{\pant}[1]{{\em #1}\xspace}

\shorttitle{Light Echoes from SN 1980K}
\shortauthors{Sugerman et al.}

\begin{document}
\title{Thirty Years of SN 1980K: Evidence for Light Echoes}

\author{Ben E.\ K.\ Sugerman\altaffilmark{1}, 
 Jennifer E.\ Andrews\altaffilmark{2},
 Michael J. Barlow\altaffilmark{3},
 Geoffrey C.\ Clayton\altaffilmark{4}, 
 Barbara Ercolano\altaffilmark{5},
 Parviz Ghavamian\altaffilmark{6},
 Robert C. Kennicutt Jr.\altaffilmark{7},
 Oliver Krause\altaffilmark{8},
 Margaret Meixner\altaffilmark{9},
 Masaaki Otsuka\altaffilmark{10}}
\altaffiltext{1}{Department of Physics \& Astronomy, Goucher College, 1021
  Dulaney Valley Rd., Baltimore, MD 21208, USA; ben.sugerman@goucher.edu}
\altaffiltext{2}{Department of Astronomy, University of Massachusetts,
  710 North Pleasant St., Amherst, MA 01003, USA}
\altaffiltext{3}{Department of Physics and Astronomy, University
  College London, Gower Street, London WC1E 6BT, UK}
\altaffiltext{4}{Department of Physics \& Astronomy, Lousiana State
  University, 202 Nicholson Hall, Baton Roughe, LA 70803, USA}
\altaffiltext{5}{Universit\"{a}ts-Sternwarte M\"{u}nchen, Scheinerstr. 1,
  81679 M\"{u}nchen, Germany}
\altaffiltext{6}{Department of Physics, Astronomy \& Geosciences,
  Towson University, Smith Hall, Towson, MD 21252, USA}
\altaffiltext{7}{Institute of Astronomy, University of Cambridge,
  Madingley Road, Cambridge, CM3 0HA, UK}
\altaffiltext{8}{Max Planck Institute for Astronomy, K\"onigstuhl 17,
  69117 Heidelberg, Germany}
\altaffiltext{9}{Space Telescope Science Institute, 3700 San Martin
  Dr., Baltimore, MD 21218, USA}
\altaffiltext{10}{Institute of Astronomy and Astrophysics, Academia
  Sinica, Taipei 10617, Taiwan, R.O.C.
}

\begin{abstract}

We report optical and mid-infrared photometry of SN 1980K between 2004
and 2010, which show slow monotonic fading consistent with previous
spectroscopic and photometric observations made 8 to 17 years after
outburst.  The slow rate-of-change over two decades suggests that
this evolution may result from scattered and thermal light echoes
off of extended circumstellar material.  We present a semi-analytic
dust radiative-transfer model that uses an empirically corrected
effective optical depth to provide a fast and robust alternative to
full Monte-Carlo radiative transfer modeling for homogenous dust at
low to intermediate optical depths.  We find that unresolved echoes
from a thin circumstellar shell 14--15 lt-yr from the progenitor, and
containing $\lesssim 0.02$ \msun of carbon-rich dust, can
explain the broadband spectral and temporal evolution.  The size, mass
and dust composition are in good agreement with the contact
discontinuity observed in scattered echoes around SN 1987A.  The origin
of slowly-changing high-velocity [\ion{O}{1}] and H$\alpha$ lines is
also considered.  We propose an origin in shocked high-velocity
metal-rich clumps of ejecta, rather than arising in the impact of
ejecta on slowly-moving circumstellar material, as is the case with
hot spots in SN 1987A.

\end{abstract}


\newpage

\section{INTRODUCTION \label{sec-intro}}

Nine supernovae (SNe) have been observed in the spiral galaxy NGC 6946
\citep[$d=5.9$ Mpc,][]{KSH00} in the last century, earning this nearby
supernova factory the nickname ``The Fireworks Galaxy'' and making it
a virtual laboratory for studying a variety of SN characteristics 
from progenitor identification through evolution into early remnant
stages.  With the {\em Hubble} and {\em Spitzer Space Telescopes}
({\em HST} and {\em Spitzer}, respectively) and
large ground-based observatories all available during
the last decade, a great deal of understandable interest has been paid
to its recent SNe 2002hh, 2004et, and 2008S.  Continued monitoring of
its older population (SNe 1917A, 1939C, 1948B, 1968D, 1969P, and
1980K) provide an extraordinary opportunity to connect the early
evolution ($\lesssim 10$ yrs) of young SNe with evolved remnants
(e.g.\ Cas A, N 103B).  In this paper, we report and analyze optical
and infrared (IR) data collected between 2004 and 2010 of SN 1980K.

First discovered on 1980 Oct 28 \citep{Wil80}, this
core-collapse, linear (or type II-L) SN was recovered in H$\alpha$
imaging in 1987 and spectroscopy in 1988 \citep{Fes90}, which revealed
a faint continuum underlying a complex of high-velocity H$\alpha$ and
[\ion{O}{1}] lines that had not been present during the first two
years \citep{Bar82,Uom86}.  Continued observations
\citep{Uom91,Fes94,Fes95,Fes99} have shown the total fluxes to have
declined by about 25\% over more than a decade, with the centers and
fluxes of the individual high-velocity components changing on
timescales of a few years. A detailed consideration of possible
mechanisms led \citet{Fes99} to conclude that these lines originate in
shocks driven by the SN blast into pre-existing circumstellar material
(CSM).  The presence of such material has been inferred both from
IR excesses observed up to one year after outburst
\citep{Dwe83b}, which \citet{Dwe83a} argued arose from thermal echoes
off of nearby CSM, and from 6 and 20 cm radio evolution \citep{Wei92}
consistent with the forward blast impacting dense CSM laid down by the
progenitor red supergiant a few tens of thousands of years prior
to core collapse.

\begin{figure*}[t!]\centering
\includegraphics[height=5in,angle=-90]{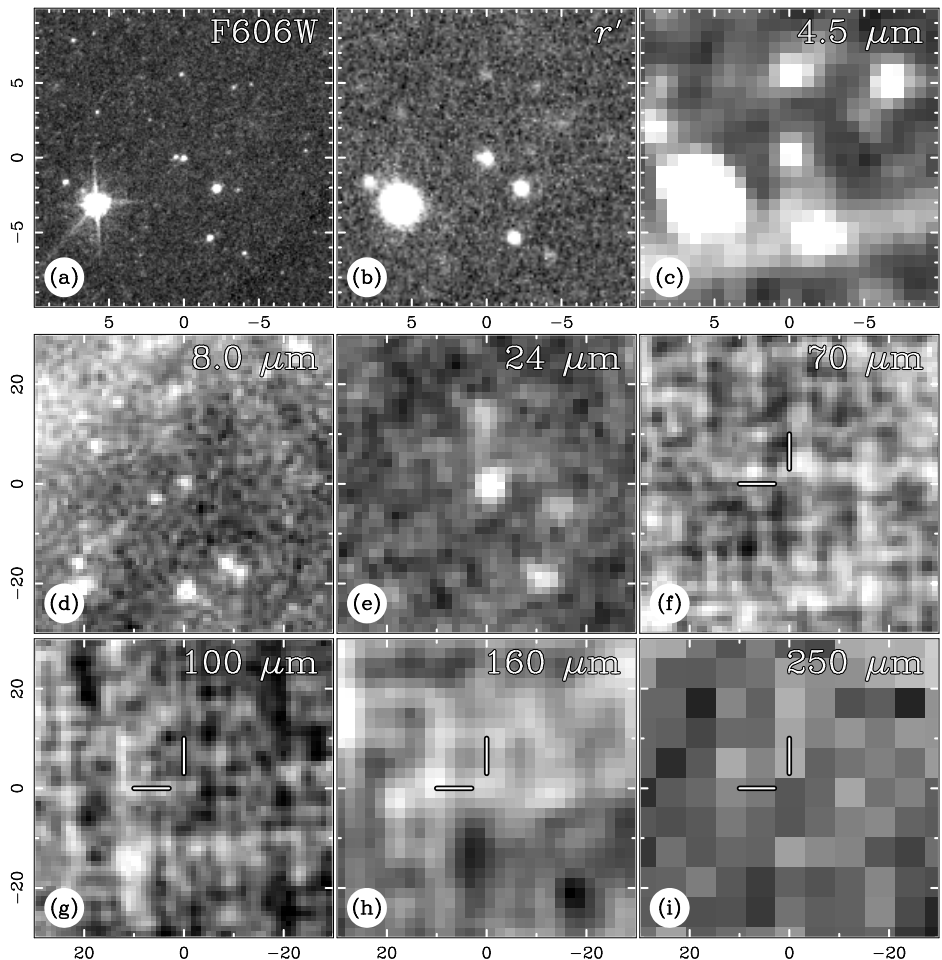}
\caption{Tiling of images of SN 1980K from optical through
  far-infrared.  Note that the image scale (arcsec) changes between
  the top and bottom two rows.  \panp{a} {\em HST}/WFPC2 F606W on 19
  Jan 2008.  \panp{b} Gemini $r'$ on 05 Aug 2005.  \panp{c} {\em
    Spitzer}/IRAC 4.5 \micron on 05 Jan 2011.  \panp{d} {\em
    Spitzer}/IRAC 8.0 \micron on 18 Jul 2008.  \panp{e} {\em
    Spitzer}/MIPS 24 \micron on 29 Jul 2008.  \panp{f-i} {\em
    Herschel}/PACS and SPIRE on 03 Oct 2010 at 70 \micron, 100 \micron,
  160 \micron, and 250 \micron, respectively.  
 \label{bigtile}}
\end{figure*}
The presence of a faint continuum within all late-time spectra, and the
slow monotonic decrease in flux in all optical wavebands, can also
arise from a scattered-light echo \citep{Cou39} illuminating a large
CSM shell or interstellar dust sheet.  Whether from scattered or
thermally-reprocessed SN light, an echo arises when dust redirects SN
light back into the line of sight, which arives at the observer later
than the original signal by the light-travel time along the longer
path length.  Indeed, a scattered echo has been considered for SN
1980K by \citet{Che86}, who concluded that the flux evolution through
the first $\sim 1$ year was consistent with a scattering-echo model,
but more observations were necessary to distinguish it from
radioactive-energy deposition.

In this paper, we present optical photometry of SN 1980K made by the
{\em HST} and Gemini North, and mid-IR photometry made by {\em
  Spitzer}, all between 2004--2010, as well as upper limits in the
far-IR from {\em Herschel Space Observatory} imaging made in 2010
\secp{sec-data}.  Combined with previously-reported optical
photometry, we show that the last decade of evolution is consistent
with a slow, monotonic fading across all wavebands \secp{sec-phot}
that can be explained by scattered and thermal echoes from a large
($r\sim 14$ lt-yr) shell of circumstellar dust.  The basic behavior
and modeling of light echoes are discussed in \sect{sec-le}, followed
by an exploration of the parameter space of possible shell properties,
as well as a new, semi-analytic dust radiative-transfer model, in
\sect{sec-tau}.  The consistency of the proposed light-echoes with
previous analyses of the high-velocity emission-line structures is
discussed in \sect{sec-comp}.

\section{OBSERVATIONS} \label{sec-data}

\subsection{Imaging} 

In the mid-infrared, SN 1980K has been imaged at least yearly between
2004 Jun and 2008 Jul with {\em Spitzer} in the
same field as SN 2004et with both the InfraRed Array Camera (IRAC) and
Multiband Imaging Photometer for Spitzer (MIPS) as part of the
original cold mission (GO-159, 20256, 20320, 30494, 40010), and
continuing through 2010 with IRAC as part of its ongoing warm mission
(GO-60071, 70008), often as part of the ongoing work of the SEEDS
collaboration (Search for Evolution of Emission from Dust in
Supernovae, P.I.\ M. Barlow).  Pipeline-calibrated imaging data were
retrieved from the {\em Spitzer} archive and processed with {\tt
  mopex} \citep{MK05} to achieve enhanced resolutions of
0\farcs75~pix$^{-1}$ for IRAC and 1\farcs5~pix$^{-1}$ for MIPS.

SN 1980K was imaged once on chip 2 of the {\em HST} Wide Field and
Planetary Camera 2 (WFPC2) in the F606W and F814W filters on 2008 Jan
19 (GO-11229).  Pipeline-reduced data were combined using {\tt
  multidrizzle} \citep{Koe02} to achive a final pixel scale of
0\farcs1~pix$^{-1}$.  Two epochs of optical observations were also
made with Gemini North in $g'r'i$ filters on 2005 Aug 05 and 2006
Jul 17 using the Gemini Multi-Object Spectrograph (GMOS) in imaging
mode.  Images were reduced and stacked using the IRAF {\tt gemini}
package, and instrumental counts were calibrated to Johnson-Cousins
$VR_cI_c$ magnitudes using transformations from \citet{Wel07}.

NGC 6946 was imaged in the far-infrared by {\em Herschel} in all
wavebands on 2010 Mar 10 as part of the KINGFISH program
(R.\ Kennicutt, P.I.).  One-arcmin fields surrounding the position of
SN 1980K were extracted from the enhanced Photoconductor Array Camera
and Spectrometer (PACS) and Spectral and Photometric Imaging REceiver
(SPIRE) data products as explained in \citet{Ken11}. Images of the
field at representative wavelengths are shown in Fig.\ \ref{bigtile}.

\subsection{Archival Spectra \label{data-spec}}

Optical spectra of various wavelength ranges and resolution have been
published for SN 1980K from maximum light in 1980 through late 1997
\citep{BCR82,Uom86,Lei91,Uom91,Fes90,Fes94,Fes95,Fes99}.  For this
work, optical spectra from the first $\sim 100$ days \citep{BCR82}
that were digitized and flux calibrated by \citet{Ben91} were acquired
from the Asiago Supernova Archive.  Spectra from \citet{Uom86},
\citet{Fes90}, and \citet{Fes99} were extracted using {\tt
  DEXTER}\footnote{http://dc.sah.uni-heidelberg.de/sdexter} from
fully-printable scans or electronic articles archived in the
Astrophysical Data System\footnote{http://adsabs.harvard.edu}.  The
late-time spectrum from 1988 Aug \citep{Fes90} was integrated over the
Johnson $V$ and Cousins $R_c$ filter response functions, and the
spectrum from 1997 Nov \citep{Fes99} over $V$, $R_c$ and $I_c$,
yielding estimates of $V=22.9$ and $R_c=22.1$ in 1988, and $V=23.3$,
$R_c=22.4$ and $I_c\lesssim 22.9$ in 1997.  These are consistent with
the broadband magnitudes $V=22.8\pm 0.2$, $R=21.9\pm 0.1$ and
$I=22.2\pm 0.3$ reported by \citet{Lei93} between 1990--1992.

\section{Photometric Analysis }\label{sec-phot}

\subsection{Astrometry \label{phot-ast}}

Given that SN 1980K is $\ge$25 years past maximum light in our data,
it is important to verify the position of the point source identified
as the SN.  15 stars in the {\em HST}/WFPC2 field were extracted from
the USNO-B1 astrometric catalog \citep{Mon03} and compared to
positions in the {\em HST} and {\em Spitzer} data to calibrate the
world-coordinate systems using standard IRAF routines.  These show
that the point source at the center of panels \panp{a--e} in
Fig.\ \ref{bigtile} are within 0\farcs1 of the VLA position of the SN
as given in \citet{Mon98}.  In particular, SN 1980K is the right-hand
point source at the very center of the {\em HST} image shown in
Fig.\ \ref{bigtile}\pant{a}.

\subsection{Photometry \label{phot-phot}}

\begin{figure}\centering
\includegraphics[height=8.7in,angle=0]{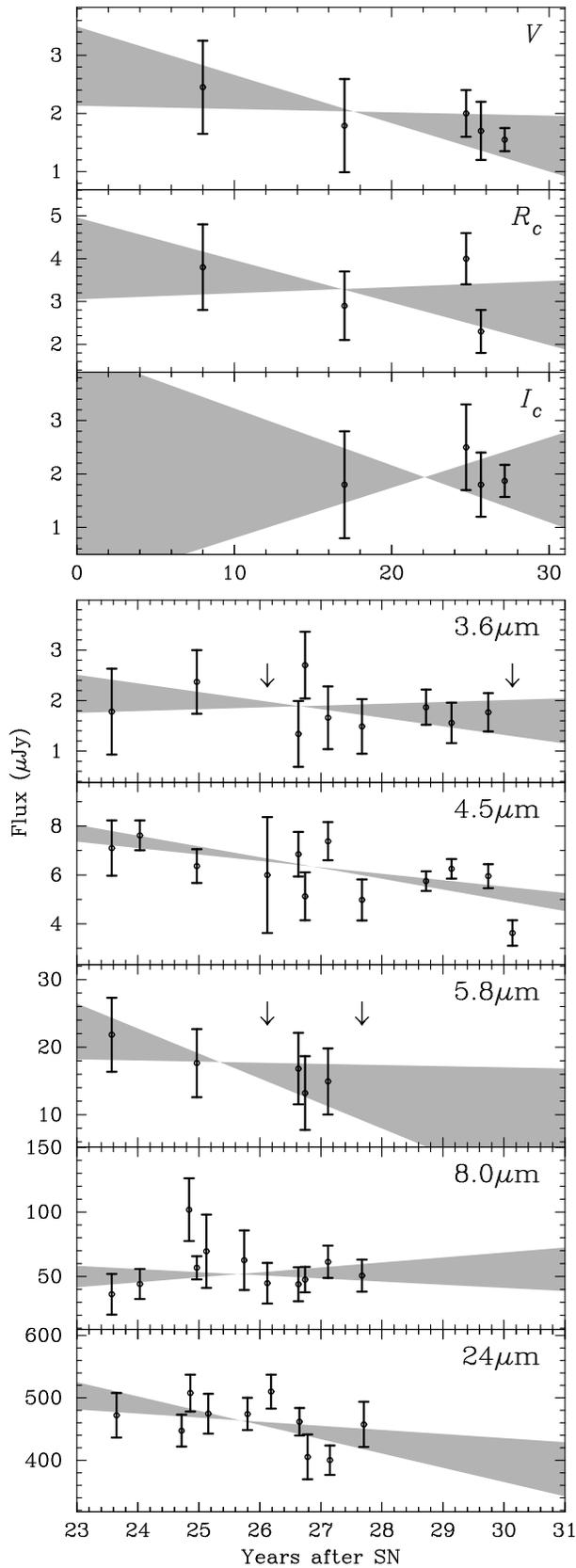}
\caption{Optical (top) and mid-IR (bottom) light curves of SN
  1980K. Wavebands are indicated in the top right of each panel. Grey
  regions indicate the range of slopes consistent with the data
  using least-squares fitting. 
 \label{multilc}}
\end{figure}

In optical and mid-IR imaging, photometry was performed using the {\tt
  daophot} PSF-fitting routines within IRAF, which also provide a
robust estimate of uncertainty using a combination of Poisson noise,
flat-fielding and absolute calibration errors.  The resulting fluxes
and errors, as well as earlier data estimated from archival spectra,
are shown in Fig.\ \ref{multilc}.  Also shown in grey are ranges of
slopes that fit the data with standard least-squares.  The general
trend is for a slow fading in time in most all wavebands, although
most data are consistent with the flux remaining approximately
constant over the time period sampled.  For this reason, the weighted
average values for each filter are presented in Table \ref{tbl-phot}
along with the least-squares slopes, and the averaged spectral-energy
distribution (SED) is plotted in Fig.\ \ref{SED}.

\begin{deluxetable*}{l c c c c c c}
\tablecaption{Photometric Behavior of SN 1980K
\label{tbl-phot}}
\tablewidth{0pt}
\tablecolumns{7}
\tablehead{
\colhead{$\lambda_0$} & \colhead{Avg.\ Flux} &
 \colhead{Best-Fit Slope} &  \colhead{$N$\tablenotemark{1}} & 
 \colhead{$Q$\tablenotemark{2}} &
 \colhead{$F(\lambda)$ at 23 yr} &  \colhead{$F(\lambda)$ at 28 yr} \\
 \colhead{($\mu$m)} & \colhead{($\mu$Jy)} & \colhead{($\mu$Jy yr$^{-1}$)} & 
 \colhead{ } & \colhead{ } & \colhead{($\mu$Jy)} & \colhead{($\mu$Jy)} 
}
\startdata
0.55  & $1.7 \pm 0.4$  & -0.044 $\pm$ 0.039 & 5 & 0.873 & 1.8$\pm$0.4 & 1.6$\pm$0.4 \\
0.64  & $3.1 \pm 0.7$  & -0.043 $\pm$ 0.057 & 4 & 0.091 & 3.0$\pm$0.7 & 2.8$\pm$0.7 \\
0.80  & $1.9 \pm 0.5$  & -0.006 $\pm$ 0.100 & 4 & 0.740 & 1.9$\pm$0.7 & 1.9$\pm$0.7 \\
3.6  & $1.8 \pm 0.5$   & -0.067 $\pm$ 0.103 & 9 & 0.831 & 2.1$\pm$0.7 & 1.8$\pm$0.7 \\
4.5  & $5.9 \pm 0.6$   & -0.352 $\pm$ 0.090 & 12& 0.024 & 7.7$\pm$0.7 & 5.9$\pm$0.8 \\
5.8  & $16.8 \pm 5.2$  & -1.936 $\pm$ 1.769 & 5 & 0.968 & 22$\pm$10  &  13$\pm$10 \\
8.0  & $51.9 \pm 13.4$ & 0.706 $\pm$ 3.147 & 11 & 0.552 & 50$\pm$21  &  53$\pm$21 \\
24.  & $460. \pm 28.$  & -14.73 $\pm$ 8.15 & 10 & 0.138 & 503$\pm$50  & 430$\pm$50 \\
70.  & $<5900. $ & \ldots & 1&  \ldots &  $<5900. $ & $<5900. $      \\
100.  & $<5900.$ & \ldots & 1&  \ldots & $<5900. $ & $<5900. $      \\
160.  & $<9900.$ & \ldots & 1&  \ldots & $<9900.$ & $<9900.$      \\
250.  & $<15000.$ &\ldots & 1&  \ldots &  $<15000.$ & $<15000.$    
\enddata
\tablenotetext{1}{Number of epochs in fit and average}
\tablenotetext{2}{The incomplete gamma function $Q$ gives the
  goodness-of-fit probability.}
\end{deluxetable*}

No point sources were detected at the position of the SN in {\em
  Herschel} data.  Although there is a point-source like feature at
the correct position in the 160 \micron frame
(Fig.\ \ref{bigtile}\pant{h}), it has the same brightness as
horizontal and vertical banding throughout the image, suggesting it is
coincident noise.  To estimate the upper limits in each frame, a grid
of PSFs were added to each image with varying brightness until {\tt
  daophot} could detect them.  The resulting PSF fluxes are close to
the upper limits quoted for each imager by the {\em Herschel} Science
Centre\footnote{http://herschel.esac.esa.int/science\_instruments.shtml},
and have been listed in Table \ref{tbl-phot} and plotted as downward arrows in
Fig.\ \ref{SED}.

\subsection{Interpretation \label{phot-interp}}

\begin{figure}\centering
\includegraphics[height=3in,angle=-90]{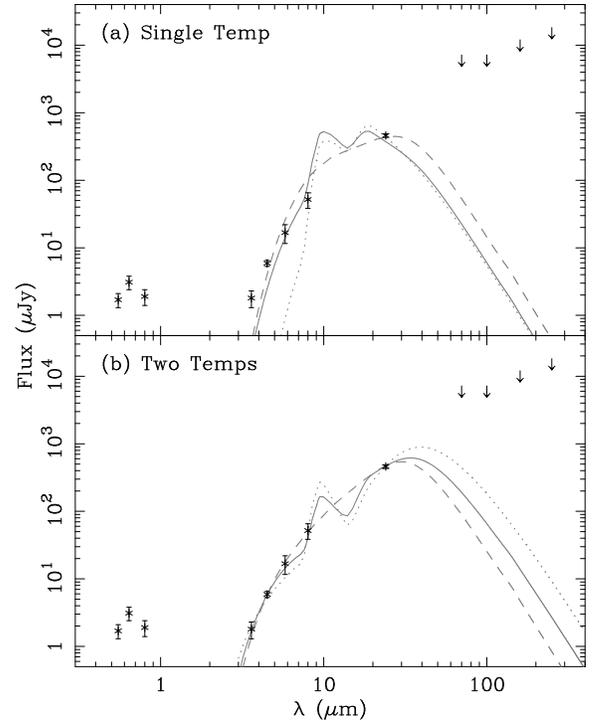}
\caption{Average SED of SN 1980K using data in Table \ref{tbl-phot}, also
  showing model spectra for simple dust emission models using \panp{a}
  one and \panp{b} two temperature components (see
  \sect{phot-interp}).  Dotted lines denote silicate dust, dashed
  lines denote graphitic dust, and solid lines show an equal mixture
  of both.  
 \label{SED}}
\end{figure}

The mid-IR SED shown in Fig.\ \ref{SED} is consistent with thermal
emission from warm dust.  Regardless of its origin, a zeroth-order
estimate of the composition and mass of emitting particles can be made
by assuming that all grains are visible (i.e. negligible internal
extinction, which is reasonable for mid-IR wavelengths), have a single
temperature $T$, and emit as blackbodies modified by their
mass-absorption coefficients $\kappa_\lambda=\pi a^2 Q_{abs} / m_g$
where $a$ is the grain radius, $Q_{abs}$ is the absorption efficiency
and $m_g$ is the grain mass.  The emerging spectrum is then given by
\begin{equation}
 F_\lambda(T) = \frac{\kappa_\lambda B_\lambda(T) M}{D^2}
\label{phot-1}
\end{equation}
where $D$ is the distance to the dust and $M$ is the total dust mass
\citep{DL94}.  Three compositions were tested using the grain
properties from \citet{WD01} and \citet{Lao93} -- pure astronomical
silicate (Si), pure graphitic solid (C), and an even mixture of both
(C+Si) -- with a standard $a^{-3.5}$ distribution of grain sizes
\citep[][hereafter MRN]{MRN77}.  Note that at mid-IR wavelengths, the
absorption coefficient $\kappa$ is relatively insensitive to grain
size.  A large range of dust temperatures were sampled for each
composition, with the dust mass at a given temperature chosen to
minimize the $\chi^2$ residual between the photometric data and
synthetic fluxes found by integrating the model SED over the {\em Spitzer}
and {\em Herschel} filter profiles.

The best-fit models are shown in Fig. \ref{SED}\pant{a}, representing
$2\times 10^{-4}$ \msun of Si dust at 200 K, $8 \times 10^{-4}$ \msun
of C dust at 230 K, and $2\times 10^{-4}$ \msun of C+Si grains at 245
K.  While the C and C+Si models come close to matching the mid-IR SED,
the fits differ from the data by $\chi^2 > 30$ suggesting that a
single-temperature model is insufficient.  We repeated the same
exercise using two dust components, where the best-fit mass for each
model was determined using the downhill-simplex algorithm {\tt amoeba}\citep{Pre92}.  For a given temperature of the warm-component, a large
range of cool-dust masses can fit the SED, due to the high upper
limits at far-IR wavelengths.  Fig.\ \ref{SED}\pant{b} shows the
best-fit models using the lowest masses for each cool-dust component.
While the C and C+Si models have $\chi^2 \sim 1$, the Si models all
have $\chi^2 > 21$, suggesting that the dust responsible for the
mid-IR emission must have some carbonaceous component.

The dust masses presented in Table \ref{tbl-2comp} represent the {\em
  instantaneous} amount of dust necessary to produce the average SED.
Since dust cooling times are short compared to the slow fading of the
light curves, we expect that either a small amount of dust is being
heated by a source with decreasing energy, or the volume of emitting
dust is changing with time.  In the former case, the dust mass present
in the system is roughly that found in this section, while in the
latter, the total dust mass is much larger than that which is emitting
at any given time.  Light echoes are one natural mechanism for
illuminating a slowly-changing volume of dust with time, where for our
timescales involved, the initial SN light pulse would be heating and
scattering off of dust in a large, pre-existing circumstellar shell.
In the next section, we present a new method for modeling such echoes,
and explore whether such echoes can reproduce the late-time data.

\begin{deluxetable*}{l c c c c c c}
\tablecaption{Results of two-component dust modeling
\label{tbl-2comp}}
\tablewidth{0pt}
\tablecolumns{7}
\tablehead{
\colhead{} & \multicolumn{2}{c}{Warm Component} &
 \multicolumn{2}{c}{Hottest Cool dust} & \multicolumn{2}{c}{Coldest
   Cool dust} \\ 
 \colhead{Model} & \colhead{$T_{warm}$ (K)} & \colhead{$M$ (\msun)} &
 \colhead{$T_{cool}$ (K)} &  \colhead{$M$ (\msun)} &
 \colhead{$T_{cool}$ (K)} &  \colhead{$M$ (\msun)}
}
\startdata
C    & $300-330$ & $4\times 10^{-5}$ & 180 & $10^{-3}$ &  55 & 3. \\
Si   & $450-480$ & $10^{-4}$ & 140 & $2 \times 10^{-3}$ & 45 & 1. \\
C+Si & $330-350$ & $10^{-5}$ & 160 &  $10^{-3}$ & 50 & 3. 
\enddata
\end{deluxetable*}

\section{Light Echoes }\label{sec-le}

\subsection{Formalism \label{le-formal}}

Light echoes occur when a pulse of light from a variable object
(e.g.\ a SN) interacts with dust and arrives at Earth some time $t$
after the pulse is directly observed.  Optical light that scatters
from that dust into the line of sight will be observed as a
scattered-light echo, while prompt thermal re-emission from dust
heated by the pulse will be observed as an infrared (or thermal) echo.
Scattered echoes from SNe have been discussed in e.g.\ \citet{Che86},
\cite{Sug03}, and \citet{Pat05} while thermal echoes have been
presented in \citet{BE80}, \citet{Dwe83a}, and \citet{EC89}.  The
following is a brief synopsis of the relevant properties of light
echoes.

An echo lies on the locus of points equidistant in light-travel from
the source and observer, i.e. an ellipsoid with the source and
observer at its foci.  In the neighborhood of the source, this
ellipsoid can be approximated by the ``echo equation''
\begin{equation}
 z^2 = \frac{\rho^2}{2ct}-\frac{ct}{2}.
\label{le-1}\end{equation}
Dust being illuminated at a time $t$ after the SN is directly
observed, and located a distance $r$ from the source at an angle
$\theta$ from the line of sight will have line-of-sight depth $z=r
\cos\theta$ and projected distance $\rho=r\sin\theta$ on the plane of
the sky.  We adopt distances and times in lt-yr and yr, respectively,
which simplifies this equation since $c=1$.  

Proceeding as in \citet{Sug03}, the flux scattered off one dust grain
of radius $a$ at position {\boldmath $r$} from the SN is 
\begin{equation} \label{le-2}
 dF_{\rm sca}(\lambda,\mbox{\boldmath $r$},a)= \frac{C_{\rm sca}(\lambda,a)
   F(\lambda)\Phi(\mu,\lambda,a)}{4\pi r^2}
\end{equation}
where $C=\pi a^2 Q$ and $Q_{\rm sca}$ is the grain scattering
efficiency, $F(\lambda)$ is the fluence (i.e.\ time-integrated flux)
at the surface of the SN, and $\Phi$ is the scattering phase function
for a given scattering angle $\mu=\cos{\theta}$.  We adopt the
\citet{HG41} phase function
\begin{equation} \label{le-3}
 \Phi(\mu,\lambda,a)=\frac{1-g^2(\lambda,a)}
     {\left[1+g^2(\lambda,a)-2g(\lambda,a)\mu \right]^{3/2}}
\end{equation}
with $g(\lambda,a)$ measuring the degree of forward scattering for a
given grain.  The total flux $F_{\rm sca}$ integrated over the
duration $\Delta t$ of the outburst from a single scattering is found
by multiplying \eqt{le-2} by the dust density $n_d(r,a)$, integrating
over the scattering volume and all grain sizes.  If the optical depth
of the cloud is low, then one can adopt the single-scattering plus
attenuation approximation \citep[SSA,][]{Pat05}, in which the flux that
reaches the dust grain is extinguished by $e^{-\tau_r}$ along the
radial path from the source to the grain, and further by $e^{-\tau_z}$
out of the dusty medium along the line of sight to the observer.
Implicit in this approximation is that no photons are scattered into
the line of sight by other grains of dust within the medium.  If the
medium consists of a distribution $f(a)$ of different grain sizes such
that $n_d(r,a)=n(r)f(a)$, the total scattered flux from a volume
element $d^3r$ arriving at Earth will be
\begin{eqnarray} \label{le-4}
\lefteqn{ F_{\rm sca}(\lambda,\mbox{\boldmath $r$}) =} \\
&& \int{ \frac{C_{\rm sca}(\lambda,a)
     F(\lambda)\Phi(\mu,\lambda,a)n_0 f(a)}{16\pi^2 D^2r^2}  
 \left(\frac{r_0}{r} \right)^p e^{-\tau_r-\tau_z}da d^3r } \nonumber
\end{eqnarray}
where $n(r)=n_0(r_0/r)^p$ allows for a power-law dependence of density with
radius, and $D$ is the distance to the SN.  Examples of grain-size
distributions include the previously-mentioned MRN
function and those of \citet{WD01}.

A grain of radius $a$ heated to temperature $T$ will radiate 
\begin{equation} \label{le-5}
 F(\lambda) = C_{\rm abs}(\lambda,a)B_\lambda(T)
\end{equation}
where $C_{\rm abs}$ is the grain absorption (and emission) cross section
and $B_\lambda(T)$ is the Planck blackbody function.  By analogy with
\eqt{le-4}, the total thermal emission from all grains within a volume element
arriving at earth will be 
\begin{equation} \label{le-6}
 F_{th}(\lambda,\mbox{\boldmath $r$})= \int{ \frac{C_{\rm abs}(\lambda,a)
     B_\lambda(T)n_0 f(a)}{4\pi D^2} \left(
   \frac{r_0}{r} \right)^p e^{-\tau_z}da d^3r }
\end{equation}

\subsection{Illuminating Spectra \label{le-spec}}

\begin{figure*}\centering
\includegraphics[height=6in,angle=-90]{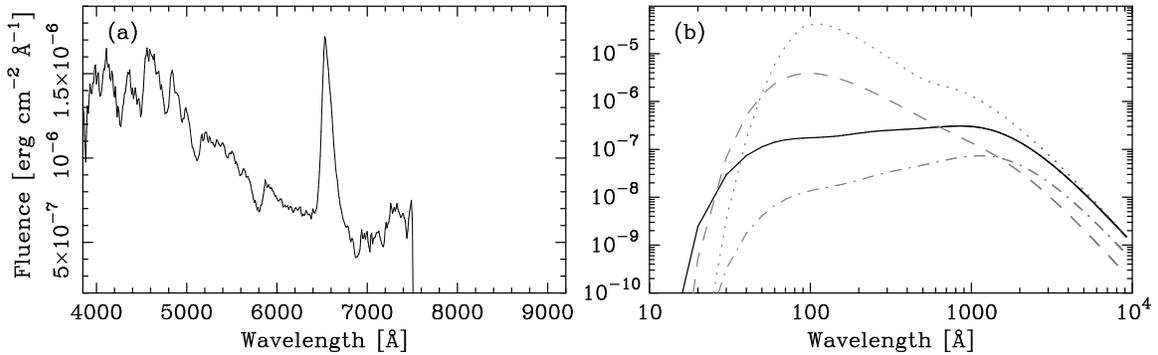}
\caption{SN Fluences used for modeling echoes from SN 1980K.  \panp{a}
  The composite optical spectrum of SN 1980K integrated over the first
  $\sim$100 days, and dereddened by $E(B-V)=0.41$. \panp{b} Different
  models for the short, prompt UV outburst.  The solid black line is
  from \citet{CF08}, while in grey, the dashed line is from
  \citet{Nak10}, dot-dashed from \citet{Rab11}, and dotted from
  \citet{Gez08}.
 \label{plibb}}
\end{figure*}

The optical spectrum $F(\lambda)$ that produces scattered-light echoes
is taken to be the time-integrated fluence over the first $\sim 100$
days, constructed by interpolating each spectrum onto a regular
wavelength grid and integrating them over time using the trapezoidal
rule.  When considering only scattered-light, one may model the echoes
using the observed (i.e.\ reddened) spectrum since it has suffered the
same interstellar extinction as any echo flux, however as we wish to
also model the thermal response to the illuminating fluence, the
dereddened spectrum must be used.  \citet{Fab11} adopt a reddening of
$E(B-V)=0.41$ toward the nearby SN 2004et, noting that the foreground
Galactic contribution is estimated to be $E(B-V)=0.34$.  If SN 1980K
is located deep within or behind the disk of NGC 6946, one may expect
it to have produced larger-radius light echoes, as have been observed
for example around SNe 1987A \citep{Xu95}, 1993J \citep{SC02}, and
2003gd \citep{S05}.  Careful inspection of all optical images show no
such echo features, thus in our modeling, we consider that while SN
1980K may suffer the same extinction as SN 2004et, it may also lie on
the near-side of its host galaxy and its extinction may be closer to
the lower, Galactic value.  In what follows, we designate these the
``high'' and ``low'' extinction models, respectively.  The fluence
spectrum, dereddened by the full $E(B-V)=0.41$ is shown in
Fig.\ \ref{plibb}\pant{a}.  Note that since this fluence spectrum ends
at 7500 \AA, it is not possible to model the $I$-band scattered flux
of any echoes.

Modeling the thermal echo of a SN requires considering all the photon
heating sources of the dust, which includes not only the long-duration
optical pulse but the preceding short-duration but highly-energetic UV
pulse, which has been variously modeled by \citet{Gez08},
\citet{CF08}, \citet{Nak10}, and \citet{Rab11}.  The blackbody
parameters presented in each paper were used to create fluence spectra
over the first $\sim 2.3$ days, as shown in Fig.\ \ref{plibb}\pant{b}.
Although the \citet{CF08} model was proposed for Ib/c SNe, its fluence
spectrum describes a reasonable average between the other three models
considered, therefore we adopt this spectrum as describing the initial
UV pulse.

\subsection{Constraints \label{le-constraint}}

Fig.\ \ref{bigtile}\pant{a} shows that there is no resolved flux
around SN 1980K, and examination of the central source shows that it
is consistent with a point source, thus any echo must come from
material such that its flux remains unresolved.  Experimentation with
marginally-resolved artificial sources show that an echo can remain
unresolved with a projected distance $\rho$ up to 1.2 times the
full-width at half-maximum (FWHM) of the PSF, which for the WF2 chip
yields $\rho\lesssim 0.2\farcs$ or roughly 19 lt-yr.  At $t=28$ yrs,
an echo would be at $z\sim-7$ lt-yr and $r\sim 20$ lt-yr.  For an echo
to be present $t$ yrs after outburst, material must be at least $t/2$
lt-yr from the source.  Since there is no other spatial information
available, we will model the echo as arising from a simple spherical
shell with outer radius between 14 and 20 lt-yr.  \citet{CE89} argue
that red supergiant winds expanding into ambient interstellar media
will have radius of order 16 lt-yr, consistent with these constraints.

Since the average light curves presented in Fig.\ \ref{multilc} are
monotonically fading in almost all wavebands, we opt to model the
system using the average optical and mid-IR data from 23 and 28 yrs
after outburst, which cover the period of multi-waveband {\em Spitzer}
observations.  Fluxes and their associated uncertainties for each
epoch were taken as the least-squares fits to each monochromatic light
curve, and are listed in the two rightmost columns in Table
\ref{tbl-phot}.  Lacking any temporal data in the far-IR, we assume
that the upper limits from {\em Herschel} apply throughout the epochs
studied.

\subsection{Modeling Echoes \label{le-model}}

The optical and thermal echo SED resulting from a SN will be the sum
of the light scattered from the short ($\sim2$ day) UV and long ($\sim
150$ day) optical pulses, as well as the thermal emission resulting
from the dust being heated by those two input spectra.  The UV pulse
contributes negligibly to the optical echo, however both spectra are
important sources of heating for thermal emission. 

The equilibrium
temperature of dust illuminated by spectrum $F(\lambda)$ (and
excluding collisional heating, which is reasonable in low-density,
slow-moving, unshocked circumstellar media) is given by
\citep[e.g.][]{RR80}
\begin{equation}\label{le-7}
4\int{C_{\rm abs}(\lambda,a)B_\lambda(T) d\lambda} =
\frac{1}{4\pi r^2}
 \int{F(\lambda)C_{\rm abs}(\lambda,a)e^{-\tau_r}d\lambda}.
\end{equation}
Numerically, the dust temperature for a given grain size $a$ can be
calculated using a root-finding algorithm such as Brent's method
\citep{Pre92} which varies $T$ until both sides of \eqt{le-6} are
equal.  If dust is not heated beyond $\lesssim 1000$ K, the majority
of its emission will be in the mid-IR and will contribute
insignificantly to re-heating of neighboring dust.  A thermal echo
spectrum can thus be computed as the emission coming from the adjacent
parabolic regions illuminated by the UV and optical pulses at a given
time $t$ after outburst by solving \eqt{le-7} at each point along the
echo parabola and summing the contributions using
\eqt{le-6}. Similarly, the scattered optical component is computed by
summing \eqt{le-2} in the same regions.  

\begin{figure*}\centering
\includegraphics[height=6in,angle=-90]{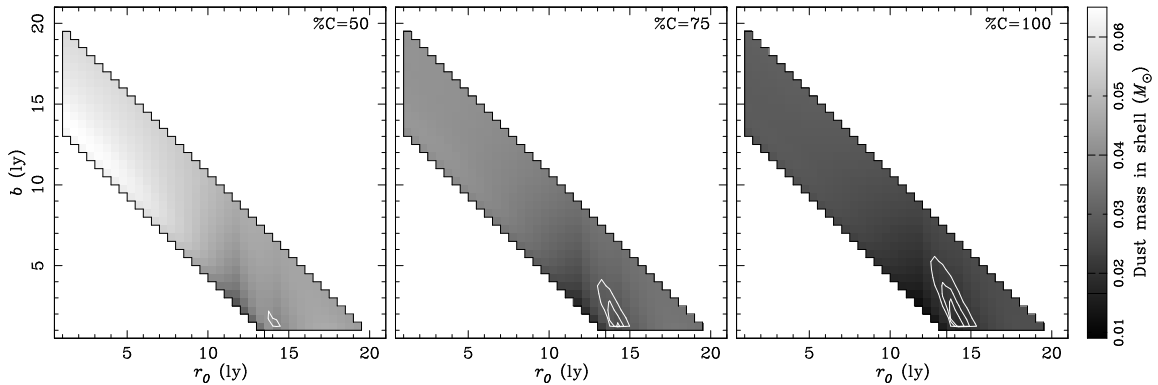}
\caption{Results of QuickSAND modeling of optical and thermal echoes
  to reproduce the average photometry of SN 1980K at 23 and 28 yrs
  after outburst (Table \ref{tbl-phot}).  Each panel shows in
  greyscale the dust mass that best fits the data for a constant
  density ($p=0$), low-extinction model with the given values of
  inner-shell radius $r_0$, shell thickness $b$ and carbonaceous-dust
  fraction $c$.  The wedge at right shows the scaling between shading
  and the shell's dust-mass.  Overplotted on each panel are contours
  of constant reduced $\chi^2$ at intervals of $\chi_{red}^2=0.5$,
  with the outermost tracing $\chi_{red}^2=2$.  The results for high
  extinction are nearly indistinguishable, with the contour interval
  changed to unity and the outermost contour tracing
  $\chi_{red}^2=4$.  In the neighborhood of the good fits, the $p=2$
  models are also nearly indistinguishable.  
 \label{cont_mass}}
\end{figure*}

The analytic expressions presented above are only appropriate in the
optically-thin regime, since multiple scatterings within an
optically-thick medium return photons to a given path as well as
remove them.  \citet{Pat05} has shown that for a variety of dust
geometries, the SSA breaks down for optical depths $\tau\gtrsim 0.1$
and that a proper treatment requires Monte Carlo simulations.  A
variety of dust radiative-transfer Monte-Carlo models exist, however
they are numerically expensive, with the result that exploration of a
large parameter space of dust geometries and densities to fit a given
dataset can require tens of thousands of computing hours.  A further
complication is that of resolution: most three-dimensional Monte-Carlo
models track photon paths through a spatial grid of cells, making the
run time and memory allocation dependent on $N^3$ where $N$ is the
number of cells in a given dimension.  The UV pulse lasts only
$\sim 2$ days, however a star's CSM can span tens of lt-yr.  Without
an adaptive mesh-refinement scheme, the spatial sampling needed to
describe the dust's behavior in this narrow echo requires a
prohibitively high numbers of grid cells.

It is for these reasons that we have developed the QuickSAND
semi-analytical model presented in Appendix \ref{sec-tau}, which
approximates the results of Monte-Carlo dust radiative-transfer models
in homogenous media to a high degree of accuracy.  For an alternative
semi-analytic model, please see \citet[][hereafter VD99]{VD99}.

To test whether the proposed scattered and thermal echoes can
reproduce the broadband photometry of SN 1980K, we ran the QuickSAND
model on a large parameter space of homogenous spherical shells, where
each possible shell is described by an inner radius $r_0$, 
thickness $b$, density power-law index $p$ \secp{le-formal}, and
carbonaceous-dust fraction $c$ (such that the amount of silicate dust
is $1-c$).  Values of $r_0$ and $b$ were tested every 0.5 lt-yr such
that $r0+b<20$ lt-yr \secp{le-constraint}, with $p=0$ and 2, and
$c=\frac{1}{2}$, $\frac{3}{4}$, and 1 (recall modeling from
\sect{phot-interp} showing silicate-dominated dust does not reproduce
the SEDs).  As noted in \citet{Sug03}, one can use the {\em observed}
spectrum when modeling a scattered-light echo since the observed
spectrum has already suffered the appropriate extinction along the
line-of-sight to the observer.  However, the extinguished spectrum is
inappropriate for heating of dust as the dust receives the
unextinguished light.  The QuickSAND SEDs were constructed using the
unextinguished optical and UV bursts in Fig. \ref{plibb}, and then
reddened by $E(B-V)=0.42$ to compare against data.

Adopting the same grains as were used in \sect{phot-interp}, we used
Brent's method to determine the dust density that minimizes the
$\chi^2$ residuals between the QuickSAND model SEDs and data at both
23 and 28 yrs (Table \ref{tbl-phot}) for each set of input parameters.
This is another advantage of our semi-analytic model over full
radiative-transfer codes, since QuickSAND can automatically find the
best-fit density rather than the user having to modify it by hand
through trial and error.  Another option is to allow such
models to find the best fit of all parameters, by using a
multi-dimensional minimization scheme such as the downhill simplex,
however there are many potential problems with this approach, namely,
the minimization routine can settle into local minima, the output does
not provide uncertanties or ranges of parameter space that fit data,
and the runtime is significantly longer than stepping through a grid
of parameters while minimizing only one (e.g. density).

The results for the low-extinction model are summarized graphically in
Fig.\ \ref{cont_mass}, which shows in greyscale the dust mass that
best fits the data for differing values of $r_0$, $b$, and dust
composition, assuming constant density with radius ($p=0$).
Overplotted in white are contours of reduced $\chi^2$ (i.e. $\chi^2$
per degree of freedom) at intervals of $\chi_{red}^2=0.5$ with the
outermost countour tracing $\chi_{red}^2=2$.  The best-fit models are
thin shells with $r_0\sim 14$ lt-yr, $b=1-4$ lt-yr, a medium dominated
by carbonaceous dust, and a total dust mass of $0.02-0.05$ \msun.  As
normalized by \citet{WD01}, the MRN grain distribution corresponds to
a gas-to-dust mass ratio $\sim 150$, which yields a total shell mass
$\sim 3$ \msun.

\begin{figure}\centering
\includegraphics[width=4in,angle=-90]{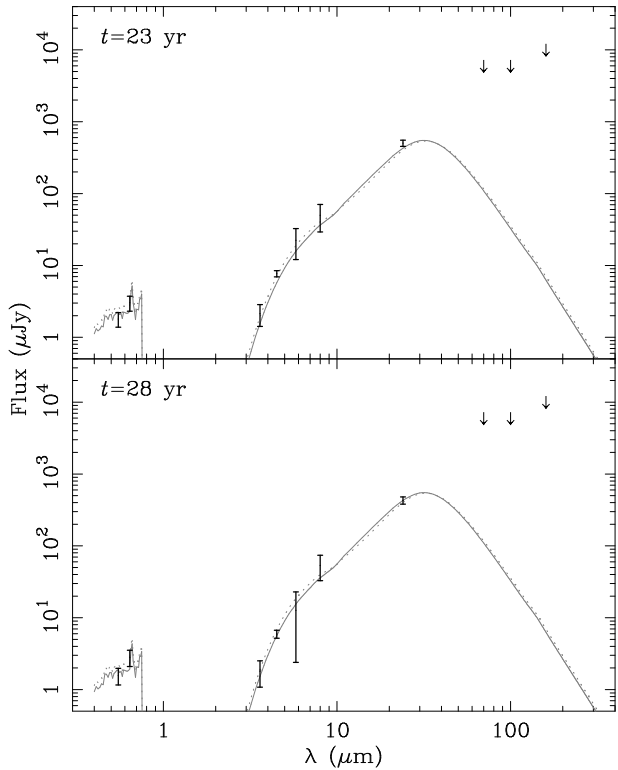}
\caption{Best-fit high (solid) and low-extinction (dotted) QuickSAND
  models of the average optical and mid-IR fluxes at 23 ({\em top})
  and 28 ({\em bottom}) years after outburst.  Both models are for a
  shell with $r_0$=14 lt-yr, $b$=1 lt-yr and $c=1.0$, with
  $M_{d}=1.5\times 10^{-2}$ \msun and $\chi^2_{red}=1.6$ in the
  high-extinction case, and $M_{d}=1.8\times 10^{-2}$ \msun and
  $\chi^2_{red}=1.$ for low extinction.   
 \label{echosed}}
\end{figure}

Results for the higher-extinction model are nearly indistinguishable,
except the contours are spaced at intervals of $\chi_{red}^2=1.$ with
the outermost contour tracing $\chi_{red}^2=4$.  For both low and high
extinction cases, the $p=2$ models also yield nearly indistinguishable
results in the neighborhood of the good fits.  In all models, the
$V$-band optical depths of the good fits are $\lesssim 10^{-3}$,
i.e.\ the proposed CSM is optically thin.  The SEDs for the best-fit
low and high-extintction models are plotted against the data in
Fig.\ \ref{echosed}.

Such a structure may be the contact discontinuity between
the star's winds and the ambient ISM (also sometimes called a
wind-blown bubble).  The size scale, mass and dust-grain properties
are roughly consistent with the contact discontinuity around SN 1987A
reconstructed by \citet{Sug05}, which was found to be a pinched
prolate spheroid, with polar and equatorial axes of 20 and 11 lt-yr,
shell thickness $\sim 4$ lt-yr, and total mass $\gtrsim 1.2$ \msun,
assuming a gas-to-dust ratio $\sim 500$.  Recall that SN 1987A is also
surrounded by a complex CSM consisting of three illuminated rings and
a dusty peanut-shaped structure, located $\sim 1-2$ lt-yr from the
progenitor.  If SN 1987A is indicative of the mass-loss structures
expected around high-mass stars, then a large wind-blown bubble
(representing main-sequence and/or early red-supergiant winds) is
expected to be accompanied by inner circumstellar structures that were
formed shortly before core collapse.  Indeed, \citet{Dwe83a} invoked
thermal echoes from such inner material to explain early-time IR
excesses from SN 1980K.

The deduced structure could also be consistent with an evolved version
of the type of shell found around Luminous Blue Variables (LBVs) such
as AG Car, He~3-591 and Sk $-69$\degr~279, with respective shell
dimensions of 6.5$\times$4.5~lt-yr, 1.6$\times$1.6~lt-yr, and
20.2$\times$14.7~lt-yr. \citep{Wei11}.  The nebulae around the
massive central stars of the AG~Car and He~3-591 nebulae have been
estimated to have cool dust masses of 0.25 \msun and \msun,
respectively \citep{Voo00}. The mass loss rate of AG~Car has
been estimated to range between 1.5--3.7$\times10^{-5}$
\msun~yr$^{-1}$ \citep{Bar91,Gro09}, comparable to the
mass loss rate estimated for the immediate progenitor of SN~1980k by
\citet{Dwe83a}. The mid-IR spectrum of AG Car is featureless apart from a
weak 11.3 \micron PAH emission band (Voors et al. 2000, Rajagopal et
al. 2007), also consistent with the dust properties found here to be
needed to match the spectrum of the light echo emission around
SN~1980k.   Regrettably, the observations of and
constraints from all echoes are insufficient to warrant a more
thorough morphological investigation of these proposed inner and outer
circumstellar echoes.

\begin{figure}\centering
\includegraphics[height=3.3in,angle=-90]{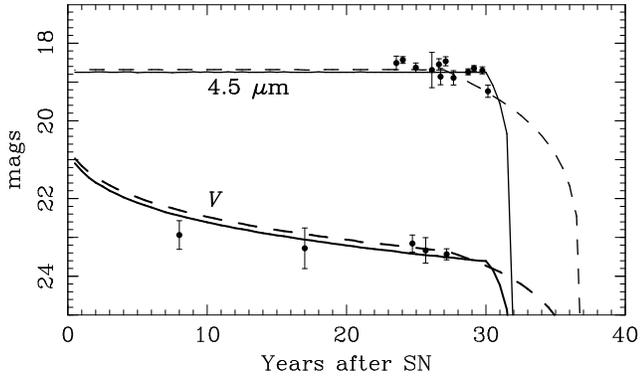}
\caption{Light curves in $V$ (bottom) and at 4.5 \micron (top) of
  echoes off of thin (solid) and thick (dashed) spherical shells that
  fit the observed data, along with the observed data in both
  wavebands.  
 \label{echolc}}
\end{figure}

As discussed in \sect{le-constraint}, an echo will pass out of a
spherical shell of size $t/2$ lt-yr in $t$ years, thus it is important
to note that the best-fit models imply that the echo left the CSM
shortly after $t=30$ yrs.  Fig.\ \ref{echolc} shows the expected $V$
and 4.5 \micron light curves for thin ($r0=15$, $b=1$) and thick
($r0=13.5$, $b=5$) shells that are consistent with the data. Note the
precipitous drop in flux in the thin shell, as compared to the slow
fading in thicker shells.  Taken on its own, the drop observed in the
4.5 \micron datum from $t=30.1$ yrs (see also Fig.\ \ref{multilc}) is
consistent with the light curve expected from the thin-shell model.
It should be emphasized that such a drop in the light curve after 30
yrs was not fit during our modeling.  Obviously, future observations
in the optical and mid-IR are required to ensure the SN is actually
fading; the epoch of this fading and its duration will serve to
disentangle the broad range of acceptable models in
Fig.\ \ref{cont_mass}.  Still, with the data in hand it appears that
light echoes from a thin shell with inner radius around 14--15 lt-yr are
an excellent explanation for the multiband observations of SN 1980K.  

\section{Comparison with Previous Analyses }\label{sec-comp}

\begin{figure}\centering
\includegraphics[width=3.3in,angle=0]{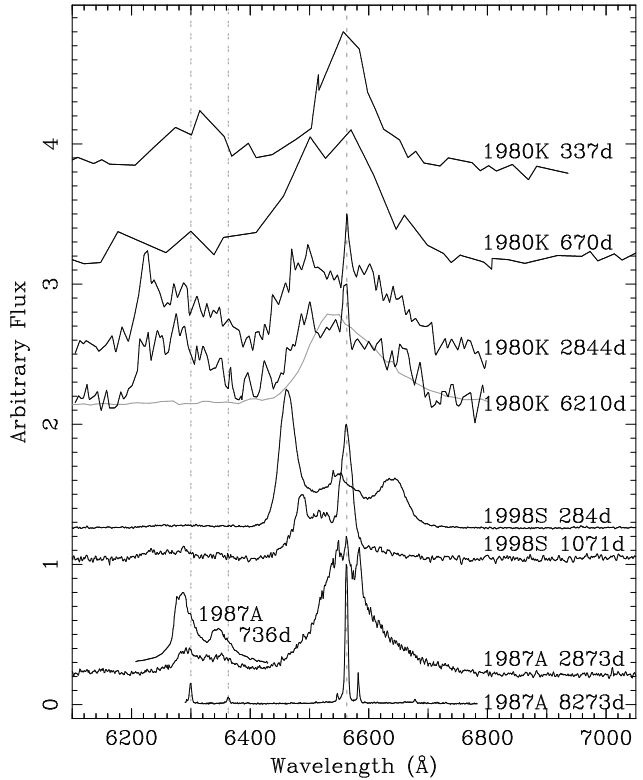}
\caption{Spectra of SN 1980K (top four), SN 1998S (middle two), and SN
  1987A (bottom three) at the dates indicated, and scaled to arbitrary
  flux.  In grey under the SN 1980K spectrum at day 6210 is plotted
  the estimated fluence of the SN from Fig.\ \ref{plibb}\pant{a} as it
  would appear in an echo when scattered by MRN dust.  The dashed thin
  grey lines show the rest wavelengths for H$\alpha$ and the
  [\ion{O}{1}] $\lambda\lambda$6300,6363 doublet.
 \label{plOI}}
\end{figure}

As noted in \S1, \citet{Fes90}, \citet{Fes94}, and \citet{Fes95,Fes99}
have reported a slowly-varying complex of highly-blueshifted
($v_r\lesssim 7000$ km~s${-1}$) H$\alpha$ and [\ion{O}{1}]
$\lambda\lambda$6300,6363 emission lines.  Representative spectra are
reproduced at the top of Fig.\ \ref{plOI}.  \citet{Fes90} ruled out
these high-velocity features as being due to a light echo, since such
features were not seen in early spectra.  Instead, they proposed that
such features might arise from ejecta shocking pre-existing
circumsteller material.  It is therefore important to consider whether
the observational signatures of such an impact are consistent with our
proposed light-echo scenario in explaining the last few decades of
evolution.

The only very-young supernova remnant that has been visually confirmed
to have ejecta-CSM impact is SN 1987A, in which the forward blast has
been lighting up ``hot spots'' around the inner equatorial ring or ``ER''
\citep[e.g.][]{Sug02,Fra10}.  Representative spectra from roughly the
same epochs as the very-late-time spectra from SN 1980K are shown at
the bottom of Fig.\ \ref{plOI}.  The day 2873 spectrum is an {\em HST}
Cosmic-Origins Spectrograph observation of the entire inner ring,
first hot-spot and ejecta taken on 1995 Jan 08, and the day 8723
spectrum is of the ring and a few bright hot spots taken with the
Space Telescope Imaging Spectrograph on 2009 Oct 21.  Both spectra
were extracted from pipeline-calibrated data taken from the {\em HST}
archive. Note the narrow H$\alpha$, [\ion{N}{2}] and [\ion{O}{1}]
lines arising from the unshocked ring material.  Whether including the
entire ring and impact system (day 2873), or only a subset (day 8273),
these spectra do not show the very-high blueshift emission lines seen in
SN 1980K.  

In the plane-parallel approximation \citep{Sgro75}, the transmitted
shock velocity $v_t$ when impacting media with overdensity $\delta$ is
\begin{equation}
v_t = v_e\sqrt{\frac{\beta}{\delta}}
\label{comp-0}\end{equation}
where $v_e$ is the ejecta
shock speed and $\beta=6(1+1.701\delta^{-1/2}-0.4018\delta^{-1}+0.2274\delta^{-3/2}-0.0874\delta^{-2})^{-2}$
\citep{BBM97b} measures the pressure differential
pre- and post-shock.  While the SN 1987A ejecta are
moving at $\sim 3500$ km~s$^{-1}$, the shocks transmitted into the ER
travel at significantly lower speeds ($\sim 200$ km~s$^{-1}$),
producing lines that are far narrower and less doppler-shifted
\citep{Mic00,Pun02} than those in SN 1980K.  It is therefore unclear
whether the high-velocity features in SN 1980K can be adequately
explained by an impact with slow-moving CSM.

Nonetheless, if this impact is indeed underway, then dust present
within the CSM will be heated either collisionally or radiatively
within the shocked and/or post-shocked gas.  \citet{Bou06} have
investigated these mechanisms for thermal emission from the hot spots
in SN 1987A, and find that the dust is most-likely heated by radiation
from post-shocked gas, has a density $\sim 10^{-8}$ cm$^{-3}$, and is
heated to $\sim 150$ K.  The SN 1980K CSM models from \citet{Dwe83a}
yield dust and gas densities similar to those in the ER of SN 1987A,
while spectral analyses from e.g.\ \citet{Fes99} suggest the ejecta
could be expanding up to $\sim 5000$ km~s$^{-1}$.  Since post-shock
gas temperature scales as $v_t^2$ (see \eqt{comp-1} below), dust that
is radiatively heated by that gas will have temperature $T_d \propto
\sqrt{v_t}$ \citep{BBM97b,Bou06}.  We can therefore approximate the
ejecta-CSM impact as heating dust to $\lesssim 300$ K, which is close
to the range of warm dust found in our two-temperature fits in
\sect{phot-interp}.  A single ``skin'' of 300 K MRN dust that is
$10^{16}$ cm thick at the $\sim 3\times 10^{17}$ cm radius of the
\citet{Dwe83a} model emits up to an order of magnitude more flux at
8 \micron than has been observed.  If one adopts the two-shock model
from \citet{Pun02}, in which slow and fast shocks are transmitted into
the CSM based on the geometry and density of the impacted medium, and
if one makes the reasonable assumption that the CSM is not a complete
spheroid (i.e. thermal emission is not coming from all 4$\pi$ sr) then
an ejecta impact could explain the average mid-IR SED, provided the
slow shock component has a velocity $\lesssim 25$\% that of the fast
shocks.

A complete investigation of the radiation expected from ejecta-CSM
impact requires time-dependent gas and dust radiative-transfer shock
modeling, which is outside the scope of this paper.  However we can
compare the heating and cooling timescales to estimate the overall
energetic evolution of such a system.  Again taking SN 1987A as our
example, \citet{Pun02} found that a forward blast traveling at $\sim
3500$ km~s$^{-1}$ from an \ion{H}{2} region with $n\sim 150$ cm$^{-3}$
into CSM with $n\gtrsim 10^4$ cm$^{-3}$ will produce shocks 
traveling at $v_t\sim 250$ km~s$^{-1}$ and heating post-shock gas to
an equilibrium temperature of $\sim 10^6$ K.  Using their cooling-time
approximation
\begin{equation}
 t_{cool} = 2.2\left(\frac{2\times 10^4\textrm{\ amu\ cm}^{-3}}{\mu_A n_0} \right)
 \left(\frac{v_t}{250\mathrm{\ km\ s}^{-1}} \right) ^{3.8}\mathrm{yr}
\label{comp-1}
\end{equation}
where $n_0$ is the pre-shock density of the impacted material, and
$\mu_A$ is the mean atomic weight, we expect radiative cooling times
of order 1--2 yr.  With similar densities and transmitted shock
velocities (preceding paragraph) we expect the cooling timescales of
shocked CSM in SN 1980K to be of this same order.  To explain the
high-velocity lines {\em and} thermal emission (i.e. seen 10--30 yr
after outburst), the ejecta-CSM shocks would have characteristic ages
$t_{shock} \gg t_{cool}$, meaning the observed flux levels scale with
the surface area of material being shocked.  For a complete or partial
shell, this would increase with time, which is inconsistent with the
observed fading across all wavebands.  It is certainly to be expected
that the ejecta from SN 1980K are impacting inner CSM, however it
appears that this mechanism can account for (at most) a small fraction
of the observed mid-IR flux, and may be predominantly responsible for
the presence of the narrow systemic emission lines.

The shape and evolution of H$\alpha$ emission lines from SN 1998S are
also indicative of ejecta-CSM interaction \citep{Poz04}; two
representative spectra are shown in the middle of Fig.\ \ref{plOI}.
Note the early appearance of a triple-peaked H$\alpha$ profile,
believed to arise from impact of ejecta with an edge-on ring
\citep{Ger00}.  By day 658, the red wing of this profile had
disappeared, likely due to preferential extinction of emission from
receding ejecta by dust that had condensed within it.  This affect was
also seen in the [\ion{O}{1}] lines in SN 1987A by \citep{Luc89}, the
spectrum of which is shown for day 736 in Fig.\ \ref{plOI}. At later
times, the emergence of a strong, central peak, slowing of the
blueshifted line, and lack of any red emission can be interpreted by
``clumpy wind'' model of \citet{CD94} in which 
intermediate-width spectral features seen at late times arise from
shocked clumps of dense wind.

Although SN 1998S is a different Type (IIn) of SN, The H$\alpha$
profile in its late spectrum bears enough resemblance to those of the
H$\alpha$ and [\ion{O}{1}] lines in the very-late spectra of SN 1980K
to suggest that the same mechanism may be at play.  One concern with
invoking ejecta shocking a clumpy progenitor wind is that slow-moving
clumps must be accelerated to high velocities to explain the
observations, and simulations of ejecta-clump impacts to model the hot
spots in SN 1987A result in post-shock gas moving at hundreds of
km~s$^{^-1}$ rather than many thousands, as needed to match the
systemic speeds of the high-velocity emission lines.  We question
whether these high-velocity features may instead arise from reverse
shocks traveling through high-velocity and high-density clumps of
metal-rich ejecta.  These have long been predicted to form in 2-D
simulations \citep{CK78}, and recently observed in fully 3-D
simulations of asymmetric core-collapse explosions \citep{Ham10}.
Furthermore, such clumps are common in evolved remnants such as Cas A
\citep{Fes11}, N63A \citep{War03}, and G292.0+1.8 \citep{Gav05}.

As above, the time-dependent modeling of such a system is
outside the scope of this paper, however it is instructive to
considering whether emission from such clumps is expected to evolve
over the observed multi-year timescales \citep{Fes99}.  We adopt the
the homologous-expansion model of \citet{Che82}, in which the size $R$
of a homologously expanding medium scales as $R\propto
t^{(n-3)/(n-s)}$ where $n$ and $s$ are the density power-law indices
of the expanding and stationary media, respectively.  Making the typical
assumption that $n=9$ and $s=2$ for ejecta expanding into a
low-density CSM, the size of an ejecta clump will grow roughly as
$t^{6/7}$.  In the \citet{Ham10} simulation, oxygen-rich clumps (which
would produce the observed high-velocity [\ion{O}{1}] lines) have
sizes around $\sim 2\times 10^{11}$ cm and gas densities $\sim
10^{16}$ cm$^{-3}$ about 9000 s after core collapse.  At an age of $\sim
300$ yrs (roughly that of Cas A), these clumps would grow to $R\sim
3\times 10^{16}$ cm and $n\sim 100$ cm$^{-3}$, which are in excellent
agreement with size scales of $1-2\times 10^{16}$ cm and densities of
$100-300$ cm$^{-3}$ for metal-rich knots in Cas A \citep{Fes01,DS10}.

At ages of 15 and 30 yr, such clumps would have sizes $R\sim 1.5\times
10^{15}$ and $\sim 4\times10^{15}$ cm, and densities $n\sim 10^{4}$
and $\sim 2\times10^3$ cm$^{-3}$, respectively, which correspond
roughly to the epochs of spectroscopic observation of high-velocity
lines, and mid-IR photometry, respectively.  An oxygen-dominated clump
($\mu_A\sim 18$) traveling outward at thousands of km~s$^{-1}$ will
have radiative cooling times $\ll 1$ yr using \eqt{comp-1}, which is
much shorter than the crossing times of 1--4 yr for the expected range
of transmitted shock velocities.  As above, a shocked clump will emit
over its crossing time, which corresponds well with the timescales
during which the [\ion{O}{1}] lines have been observed to change.

An oxygen-rich clump with the properties above will have a total mass
$\sim 10^{-5}$ \msun.  If silicate dust within this clump forms with
an efficiency of 1\%, its thermal energy output at a few hundred K
will be over an order-of-magnitude fainter than the observed mid-IR
fluxes.  However this assumes that the entire clump is radiating, and
given the short cooling times, it is likely that the mid-IR emission
from a given clump at a given time will be significantly less.  We
cannot rule out the possibility that a number of such clumps with a
variety of different pre-shock densities and geometries could provide
the temperature range needed to fit the thermal SEDs.  As with the
ejecta/CSM impact however, one expects that the number of emitting loci
will increase with time as more clumps are shocked, in contrast to our
observations of a slow mid-IR decline.  We posit that while
high-velocity clumps may explain the observed emission-line
structures, they contribute only a small fraction of the mid-IR
emission.

Behind the day 6210 spectrum of SN 1980K (Fig.\ \ref{plOI}) we have
plotted the fluence of SN 1980K \secp{le-spec} as scattered by the
best-fit thin dust shell.  Note that this echo spectrum can account
for most of the very broad H$\alpha$ profile, as well as the faint
continua present in all the late-time spectra.  Thus we believe that
the proposed scattered and thermal echoes, in combination with
high-velocity, oxygen-rich ejecta clumps, are a reasonable explanation
for the observed photometric and spectroscopic very-late-time behavior
of SN 1980K.

\section{Conclusions }\label{sec-concl}

Below we summarize the main results of this paper.

\begin{enumerate}
\item We have observed SN 1980K in optical and mid-IR wavebands
  from 23 to 30 years after outburst.  The SN was not detected in the
  far-IR bands of {\em Herschel}.  

\item The optical light curves are consistent with fading that has
  been previously reported, and the mid-IR light curves all show the
  same trend.

\item The instantaneous dust mass needed to explain the mid-IR SEDs is
  $\gtrsim 10^{-3}$ \msun, with the presence of as much as a few \msun
  of cold dust possible given the large upper limits in the far-IR.  

\item We have presented a new method for computing the effective
  optical depth of a scattering and absorbing medium \eqp{niso4}, and an
  empirical correction that is independent of the particular grain
  species or properties \eqp{niso7}.

\item When used to characterize the line-of-sight extinction to a
  particular location within a homogenous dusty medium, these
  expressions for the effective optical depth in the equations of
  scattering \eqp{le-4} and radiative balance \eqp{le-6} yield dust
  temperatures and emerging scattered and thermal spectra that are
  correct to within $\sim 10\%$ for optically-thin, and to within a
  factor of two for optically-thick media, when compared to solutions
  using the Monte-Carlo radiative-transfer code MOCASSIN.  In
  practice, this ``QuickSAND'' runs up to a few hundred times faster
  than a traditional Monte-Carlo dust radiative-transfer code.

\item We use the QuickSAND model to compute the thermal and scattered
  echo from a SN over a large parameter space of dust composition,
  density power laws, and inner and outer shell radii, and find that
  the observed optical and mid-IR SEDs are consistent with echoes from
  a thin CSM shell of inner radius $\sim 14$ lt-yr and mass $\lesssim
  0.02$ \msun of carbon-rich dust.  This is consistent with the size
  expected for the contact discontinuity between a massive star's
  winds and the ambient ISM, and with the size, composition and mass
  of the contact discontinuity found around SN 1987A.

\item Previous observations of SN 1980K have revealed
  highly-blueshifted [\ion{O}{1}] and H$\alpha$ lines, interpreted as
  arising in impact of the SN blast with pre-existing CSM.  ``Back
  of the envelope'' considerations suggest that these lines may arise
  instead in metal rich clumps of ejecta, while the ejecta-CSM impact
  may give rise instead to narrower lines observed at their rest
  wavelengths, and that neither mechanism is likely to produce
  sufficient thermal emission to explain the observed mid-IR SEDs.  

\end{enumerate}

The quality of the light-echo modeling presented here is not exclusive
to MRN dust for a spherical shell.  For example, limiting the largest
grains to sizes $\lesssim 0.1$ \micron \citep[as suggested by
  light-echo studies of inner CSM by][]{Sug05} yields acceptable fits
with nearly the same geometry but a total dust mass of up to 0.06
\msun.  Certainly one could study an enormous range of grain-size
distributions, optical constants and shell geometries
(i.e. ellipsoids, disks, hourglasses, etc.) and we expect that many of
these would yield roughly similar fits.  Given that this system is
unresolved and sparsely sampled in the optical, we must accept
the ambiguous result that the total dust mass in the proposed
spheroidal wind-blown bubble is $\lesssim 0.1$ \msun, until another
decade of contemporaneous optical and mid-IR observations are
available.  

In reconstructing the CSM of SN 1987A, \citet{Sug05} found a total of
$\lesssim 0.02$ \msun of dust (based on the ambiguity of the
gas-to-dust mass ratio), with that budget dominated by the large
contact-discontinuity.  While close to the quasi upper-limit of
$\lesssim 0.1$ \msun of dust in the same shell around SN 1980K, both
values are under the total mass of $0.3-3$ \msun of dust in the SN
1987A remnant inferred by recent {\em Herschel} observations
\citep{Mat11}.  As shown in Table \ref{tbl-2comp}, a large amount of
cool dust can reside in the SN 1980K system and remain undetected by
{\em Herschel} at the depth of the 2010 observations.  It remains
unclear whether the {\em total} dust mass present in this remnant is
consistent with that in SN 1987A.  Significantly deeper {\em Herschel}
observations are needed of SN 1980K to place more stringent limits on
this value.

\acknowledgements

The early SN 1980K spectra used in this work are part of the
Asiago-Padova Supernova Archive.  We thank Annop Wongwathanarat and
the authors of \citet{Ham10} for generously providing the data from
their simulation.  We gratefully acknowledge our referee, Fernando
Patat, for his critical reading and insightful feedback on the
manuscript, as well as bringing the ASA spectra of SN 1980K to our
attention.  Support for B.E.K.S. for this work was provided by Hubble
Space Telescope award GO-10607.  This work is based in part on data
obtained with the {\em Spitzer Space Telescope}, which is operated by
the Jet Propulsion Laboratory, California Institute of Technology,
under a contract with NASA; with the NASA/European Space Agency {\em
  Hubble Space Telescope}, which is operated by the Association of
Universities for Research in Astronomy, Inc. (AURA), under NASA
contract NAS 5-26555; with the Gemini Observatory, which is operated
by AURA under a cooperative agreement with the NSF on behalf of the
Gemini partnership; and with {\em Herschel}, an ESA space observatory
with science instruments provided by European-led Principal
Investigator consortia and with important participation from NASA.
IRAF is distributed by the National Optical Astronomy Observatories,
which are operated by the Association of Universities for Research in
Astronomy, Inc., under cooperative agreement with the National Science
Foundation.

\appendix

\section{Semi-Analytic Dust Radiative Transfer }\label{sec-tau}

\subsection{Isotropic Scattering \label{tau-iso}}

Following the arguments in \citet{RL85}, consider a
homogenous, dusty medium, characterized by absorption and scattering
cross sections  $C_{\rm abs}$ and $C_{\rm sca}$, respectively.  Traditionally,
the optical depth $\tau$ to some distance $L$ is defined as $\tau=L/l$
where
\begin{equation}\label{iso1}
 l=\frac{1}{C_{\rm abs}+C_{\rm sca}}
\end{equation}
 is the mean-free path inside that medium.  Since $(C_{\rm abs}+C_{\rm
   sca})$ is the total extinction cross section $C_{\rm ext}$, it
 follows that the fractional absorption
\begin{equation}\label{iso2}
 \epsilon=\frac{C_{\rm abs}}{C_{\rm abs}+C_{\rm sca}},
\end{equation}
which also gives the probability that a photon path will end with
absorption.  If the physical size $L$ of the medium is large compared
to the mean-free path, then the average number $N$ of mean-free paths
that the photon will travel before being absorbed is 
\begin{equation}\label{iso3}
 N=1/\epsilon
\end{equation}
since $\epsilon$ is the probability of a given step ending in
absorption.

The path of a photon in that medium can be expressed as the vector sum
$\vec{R}=\sum_{i=1}^N{\vec{r}_i}$ of the $N$ scatters taken before
absorption (or escape).  For isotropic scattering, a photon's path is
a random walk, with the well-known results that the average final
position $\langle \vec{R} \rangle$ of many photons will be zero, while the
average displacement of the photons from their starting point is the 
RMS distance $l_*=\sqrt{\langle \vec{R}^2 \rangle}$, which expands out
to 
\begin{equation}\label{iso4}
 l_*^2 = \sum_{i=1}^N{\langle r_i^2\rangle} + 
   2\sum_{i=1}^N\sum_{j=i+1}^N{\langle \vec{r}_i \cdot \vec{r_j} \rangle}.
\end{equation}
Since the scattering is assumed to be isotropic, the first term
$\langle r_i^2\rangle \approx l^2$ while the cosines in the cross terms
average to zero, yielding the expected result $l_*=l\sqrt{N}$.
Combining this with Eqs.\ (\ref{iso1})--(\ref{iso3}) yields
$l_*^2=l^2/\epsilon$ or 
\begin{equation}\label{iso5}
 l_*=\left[C_{\rm abs}(C_{\rm abs}+C_{\rm sca})\right]^{-1/2}.
\end{equation}
The optical thickness of a medium is approximately the number of
mean-free paths required to escape, i.e. $\tau=L/l$ as defined above.
For an optically thick medium where the probability of absorption is
non trivial, one must redefine optical thickness as the ratio of the
physical size $L$ of the medium to the average
distance $l_*$ between creation and destruction of a photon.  As such,
$l_*$ is often called the ``effective mean path'' and the resulting
effective optical depth 
\begin{eqnarray}
 \tau_* &=& \frac{L}{l_*} \label{iso6} \\
        &=& \sqrt{\tau\tau_{\rm sca}} \nonumber \\
        &=& \frac{\tau}{\sqrt{N}} \label{iso7}
\end{eqnarray}
where the optical depth due to absorption, scattering, and
extinction are $\tau_{\rm abs}=LC_{\rm abs}$, $\tau_{\rm sc}=LC_{\rm sca}$, and
$\tau=\tau_{\rm abs}+\tau_{\rm sc}$, respectively.  Defining the
scattering albedo $\omega=\tau_{\rm sc}/\tau$, a little algebra yields
\begin{equation}\label{iso8}
\tau_*=\tau\sqrt{1-\omega}
\end{equation}
which is equivalent to Eq.\ (14) of VD99.  


\begin{figure}\centering
\includegraphics[width=5in,angle=0]{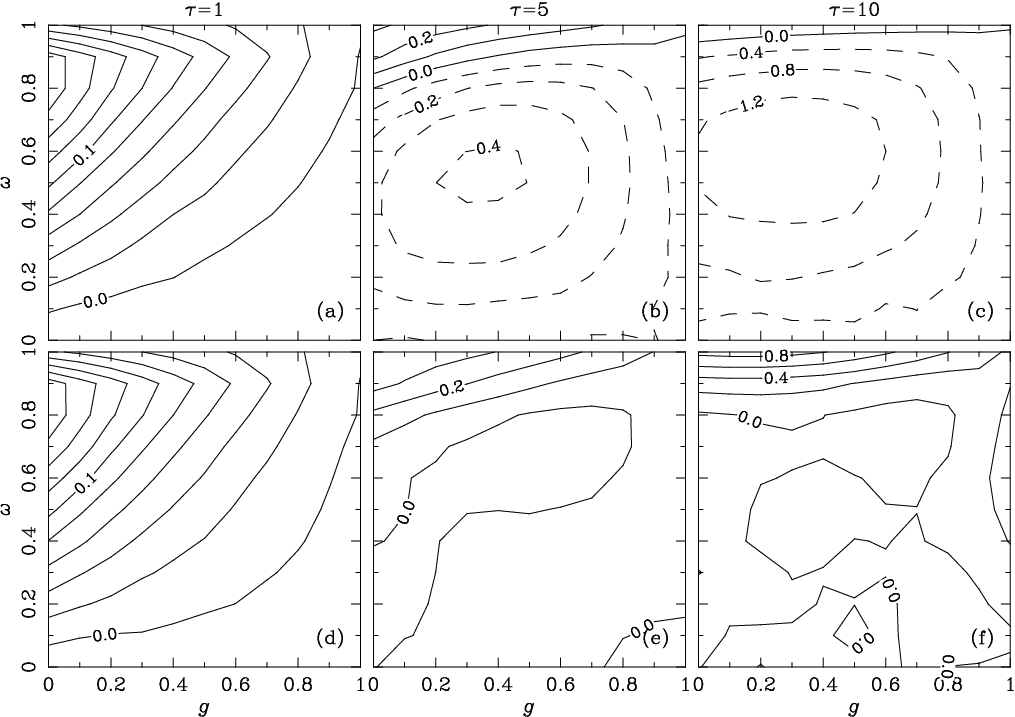}
\caption{Selected results from testing of the effective optical depth,
  showing $\tau_* - \tau_{\rm eff}$ for different values of the
  parameter triplet $\omega$, $g$ and $\tau$.  The top row computes
  $\tau_*$ with \eqt{niso4}.  The bottom row includes the empirical
  correction term $\delta \tau_*$ \eqp{niso7}.  Positive (negative)
  contours are plotted with solid (dashed) lines at the levels
  indicated.
 \label{dtau_star}}
\end{figure}

\begin{figure}\centering
\includegraphics[height=6in,angle=-90]{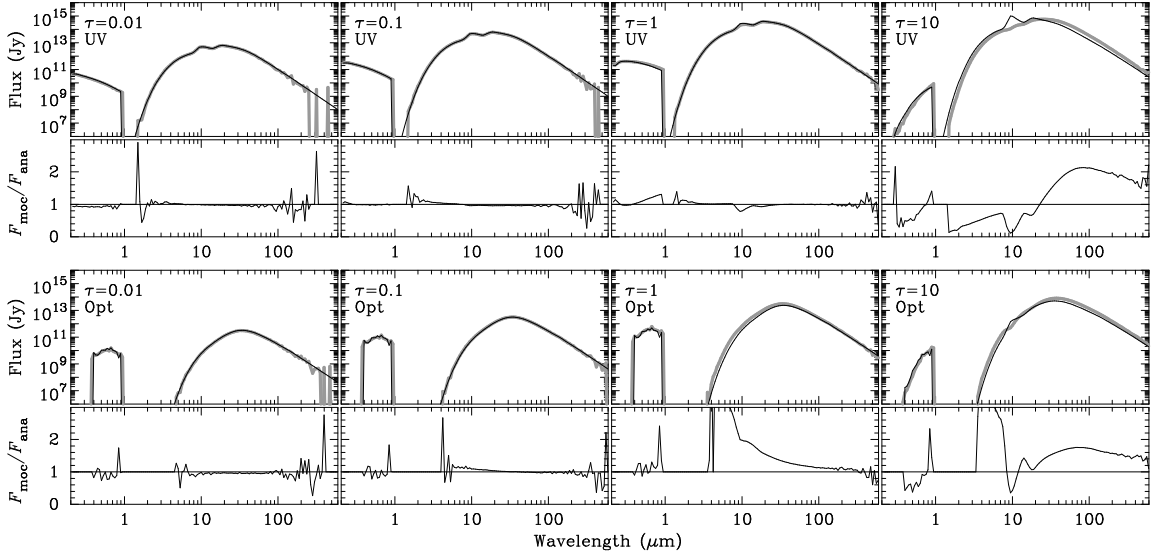}
\caption{Comparison of our semi-analytic dust radiative-transfer model
  (QuickSAND) to MOCASSIN for the SN 1980K UV (top row) and optical (bottom
  row) fluences illuminating spherical shells of optical depth listed
  at the top left corner of each panel.  The SED from MOCASSIN is
  shown in thick grey, and QuickSAND in thin black in the top half of each
  panel pair.  The bottom half of each pair shows the ratio of the two
  models.
 \label{sam_tests}}
\end{figure}

\subsection{Non-Isotropic Scattering \label{tau-niso}}

If one relaxes the assumption of isotropic scattering, a more
generalized expression can be derived by noting that the scattering
anisotropy parameter $g=\langle \cos{\theta} \rangle$, where $\theta$
is the scattering angle between incoming and outgoing photon
trajectories.  For isotropic scattering, $g=0$, while $g\rightarrow 1$
for pure forward scattering.  If $g\ne 0$, the cross terms in
 \eqt{iso4},  $\langle \vec{r}_{i} \cdot \vec{r}_{i+1} \rangle
\approx gl^2$, and by recursion, $\langle \vec{r}_i \cdot \vec{r}_j
\rangle \approx g^{j-i}l^2$.  Since $g<1$, the cross terms form a
convergent geometric sequence such that 
\begin{equation}\label{niso1}
 2\sum_{i=1}^N\sum_{j=i+1}^N{\langle \vec{r}_i \cdot \vec{r_j}
   \rangle} = 2gl^2\frac{\left( g^N-gN+N-1 \right)}{(g-1)^2}.
\end{equation}
Inserting this back into \eqt{iso4} yields
\begin{equation}\label{niso2}
 \frac{l_*}{l}=\sqrt{N_*(g)}
\end{equation}
where
\begin{equation}\label{niso3}
 N_*(g) = \frac{N-g\left( 2+gN-2g^N \right)}{(g-1)^2}
\end{equation}
and $N$ is defined as in \eqt{iso3}.
Note that the limit $N_*(g\rightarrow 1) = N^2$.
Finally, substituting \eqt{niso2} into \eqt{iso6} yields the
more general expression for effective optical depth
\begin{equation}\label{niso4}
 \tau_*(g)=\frac{\tau}{\sqrt{N_*(g)}}.  
\end{equation}
Note that \eqt{niso4} reduces to \eqt{iso7} in for isotropic
($g=0$) scattering, while for fully forward-scattering particles
($g=1$), $\tau_*(g)\approx \tau/N$ which is equivalent to Eq.\ (15) of
VD99.  

\subsection{Testing \label{tau-testing}}

We consider the dust radiative-transfer code MOCASSIN \citep{Erc05}
to be the benchmark against which the effective optical depth $\tau_*$
can be tested.  MOCASSIN self-consistently computes the
radiative-transfer of a given energy spectrum $F_0(\lambda)$ through a
dusty medium composed of user-specified grain properties within a
Cartesian grid of cells by following the paths of energy packets
(i.e.\ photons) until the emergent flux has converged to a specified
tolerance.  Since photons follow random paths, the emergent spectrum
$F(\lambda)$ has an inherent variance that depends on the number of
photons and the convergence tolerance.  

By comparing the emergent and input (or unattenuated) fluxes, the
effective optical depth of a given model is defined as
\begin{equation}\label{niso5}
 \tau_{\rm eff}=\ln{\frac{F_0(\lambda)}{F(\lambda)}}.
\end{equation}
A grid of models was constructed with $\omega$ and $g$ ranging from
$0.$ to $1.$ and optical depth $\tau$ from 0.1 to 15, allowing us to
compare $\tau_*$ to $\tau_{\rm eff}$ for each triplet of parameters.
Rather than using actual dust-scattering properties, each model used
uniform values of $C_{\rm abs}$, $C_{\rm sca}$ and $g$ for a variety of
wavelengths within a spherical shell of inner and outer radii
$10^{14}$ and $10^{15}$ cm to match the parameter triplet ($\omega$,
$\tau$, $g$) being tested.  Using a uniform input spectrum, $\tau_{\rm eff}$
is computed as the average across all wavelengths, providing a
measurement that is more robust to the inherent variance of the
Monte-Carlo process.  Selected results for $\tau=1$, 5 and 10 are
shown in Fig.\ \ref{dtau_star}.  The quality of the approximation
$\tau_*$ are roughly consistent with those presented in VD99.  The
most substantial difference is that the residuals
$\tau_*-\tau_{\rm eff}$ are well-described by an
elliptical gaussian, which is approximately modeled by the empirical
expression 
\begin{equation}\label{niso7}
 \delta\tau_*= \left\{
 \begin{array}{rl}
   (0.219\tau-0.687)e^{-4\ln{2}(\rho/0.75)^2}  & \mbox{ if $\tau\le 5.25$} \\
   7.5\times10^{-3}\tau^{2.5}e^{-4\ln{2}(\rho/0.75)^2}  & \mbox{ else}
 \end{array}
\right.
\end{equation}
where 
$\rho=\left(\frac{9}{16}(g-g_0)^2 + (\omega-\omega_0)^2\right)^{-1/2}$,
$g_0=0.3\left(1+e^{-\tau/3}\right)$ and
$\omega_0=0.585\left(1+e^{-\tau/3}\right)$.  The bottom row of
Fig.\ \ref{dtau_star} shows the result of adding this correction term
to $\tau_*$, yielding a very accurate approximation of $\tau_{\rm
  eff}$ over a very broad range of dust properties in the optically
thick regime.

\subsection{The QuickSAND Model \label{tau-sand}}

Replacing the optical depth $\tau$ in Eqs.\ (\ref{le-2}), (\ref{le-6}),
and (\ref{le-7}) with $\tau_*+\delta\tau_*$ from Eqs.\ (\ref{niso4}) and
(\ref{niso7}), we can compute the dust temperature and emergent spectra
from every point within a homogenous (i.e.\ non-clumpy) medium.  We
call this the ``QuickSAND'' (Quick Semi-ANalytic Dust)
radiative-transfer model.  

We benchmark tested the QuickSAND model against MOCASSIN by
illuminating a spherical shell of inner and outer radii of 10 and 15
ly, respectively, composed of an even mixture of carbonaceous and
silicate MRN dust \secp{phot-interp} with constant density and
illuminated by the UV and optical fluences in Fig.\ \ref{plibb}.  The
MOCASSIN models use a reasonably dense cubic grid of 61 cells in each
dimension and 128 million energy packets, both of which are needed to
ensure that sufficient photons are scattered into the line of sight to
provide adequate signal-to-noise in optical scattering and the far-IR.
The QuickSAND models sample the shell at roughly double the spatial
resolution as MOCASSIN.

Resulting benchmark SEDs from both models are shown in
Fig. \ref{sam_tests} for $\tau_V=0.01$, 0.1, 1., and 10.  The
QuickSAND model provides an excellent approximation to the MOCASSIN
model at low and intermediate optical depths, and is correct to within
a factor of two in optically-thick media.  For comparison purposes,
the MOCASSIN model running in parallel on eight processors requires
roughly 30 times longer to converge than the QuickSAND model running
on a single processor.



\clearpage


\begin{thebibliography}{}


\bibitem[Barbieri et al.(1982)]{Bar82} Barbieri, C., Bonoli, C., \&
  Cristiani, S.\ 1982, \aap, 114, 216

\bibitem[Barlow(1991)]{Bar91} Barlow, M.~J.\ 1991, IAU Symposium, 143, 281

\bibitem[Barbon et al.(1982)]{BCR82} Barbon, R., Ciatti, F., \&
  Rosino, L.\ 1982, \aap, 116, 35

\bibitem[Benetti(1991)]{Ben91}Benetti, S.\ 1991, Ph.D.~Thesis

\bibitem[Bode \& Evans(1980)]{BE80} Bode, M.~F., \& Evans, A.\ 1980,
  \mnras, 193, 21P

\bibitem[Borkowski et al.(1997)]{BBM97b} Borkowski, K.~J., Blondin,
  J.~M., \& McCray, R.\ 1997, \apj, 477, 281

\bibitem[Bouchet et al.(2006)]{Bou06} Bouchet, P., Dwek, E., Danziger,
  J., et al.\ 2006, \apj, 650, 212


\bibitem[Chevalier(1982)]{Che82} Chevalier, R.~A.\ 1982, 
\apj, 258, 790

\bibitem[Chevalier(1986)]{Che86} Chevalier, R.~A.\ 1986, \apj, 308,
  225

\bibitem[Chevalier \& Emmering(1989)]{CE89} Chevalier, R.~A., \&
  Emmering, R.~T.\ 1989, \apj, 342, 75


\bibitem[Chevalier \& Fransson(2008)]{CF08} Chevalier, R.~A., \&
  Fransson, C.\ 2008, \apjl, 683, L135

\bibitem[Chevalier \& Klein(1978)]{CK78} Chevalier, R.~A., \& Klein,
  R.~I.\ 1978, \apj, 219, 994


\bibitem[Chugai \& Danziger(1994)]{CD94} Chugai, N.~N., \& Danziger,
  I.~J.\ 1994, \mnras, 268, 173

\bibitem[Couderc(1939)]{Cou39}Couderc, P.~1939, Ann d'Ap, 2, 271

\bibitem[Docenko \& Sunyaev(2010)]{DS10} Docenko, D., \& Sunyaev,
  R.~A.\ 2010, \aap, 509, A59

\bibitem[Doty \& Leung(1994)]{DL94} Doty, S.~D., \& Leung,
  C.~M.\ 1994, \apj, 424, 729


\bibitem[Dwek(1983)]{Dwe83a} Dwek, E.\ 1983, \apj, 274, 175

\bibitem[Dwek et al.(1983)]{Dwe83b} Dwek, E., A'Hearn,
  M.~F., 
Becklin, E.~E., et al.\ 1983, \apj, 274, 168 

\bibitem[Emmering \& Chevalier(1989)]{EC89} Emmering, R.~T., \&
  Chevalier, R.~A.\ 1989, \apj, 338, 388

\bibitem[Ercolano et al.(2005)]{Erc05} Ercolano, B., Barlow, M.~J., \&
  Storey, P.~J.\ 2005, \mnras, 362, 1038

\bibitem[Fabbri et al.(2011)]{Fab11} Fabbri, J., Otsuka,
  M., Barlow, M.~J., et al.\ 2011, \mnras, 418, 1285

\bibitem[Fesen \& Becker(1990)]{Fes90} Fesen, R.~A., \& Becker,
  R.~H.\ 1990, \apj, 351, 437

\bibitem[Fesen et al.(1999)]{Fes99} Fesen, R.~A., Gerardy, C.~L.,
  Filippenko, A.~V., et al.\ 1999, \aj, 117, 725

\bibitem[Fesen et al.(1995)]{Fes95} Fesen, R.~A., Hurford, A.~P., \&
  Matonick, D.~M.\ 1995, \aj, 109, 2608

\bibitem[Fesen \& Matonick(1994)]{Fes94} Fesen, R.~A., \& Matonick,
  D.~M.\ 1994, \apj, 428, 157

\bibitem[Fesen et al.(2001)]{Fes01} Fesen, R.~A., Morse, J.~A.,
  Chevalier, R.~A., et al.\ 2001, \aj, 122, 2644

\bibitem[Fesen et al.(2011)]{Fes11} Fesen, R.~A., Zastrow, J.~A.,
  Hammell, M.~C., Shull, J.~M., \& Silvia, D.~W.\ 2011, \apj, 736, 109

\bibitem[France et al.(2010)]{Fra10} France, K., McCray, R., Heng, K.,
  et al.\ 2010, Science, 329, 1624

\bibitem[Gerardy et al.(2000)]{Ger00} Gerardy, C.~L., Fesen, R.~A.,
  H{\"o}flich, P., \& Wheeler, J.~C.\ 2000, \aj, 119, 2968

\bibitem[Gezari et al.(2008)]{Gez08} Gezari, S., Dessart, L., Basa,
  S., et al.\ 2008, \apjl, 683, L131


\bibitem[Ghavamian et al.(2005)]{Gav05} Ghavamian, P., 
Hughes, J.~P., \& Williams, T.~B.\ 2005, \apj, 635, 365 

\bibitem[Groh et al.(2009)]{Gro09} Groh, J.~H., Hillier, 
D.~J., Damineli, A., et al.\ 2009, \apj, 698, 1698 


\bibitem[Hammer et al.(2010)]{Ham10} Hammer, N.~J., Janka, 
 H.-T., Muller, E.\ 2010, \apj, 714, 1371 

\bibitem[Henyey \& Greenstein(1941)]{HG41} Henyey, L.~C.~\&
  Greenstein, J.~L.\ 1941, \apj, 93, 70


\bibitem[Karachentsev et al.(2000)]{KSH00} Karachentsev,
  I.~D., Sharina, M.~E., \& Huchtmeier, W.~K.\ 2000, \aap, 362, 544

\bibitem[Kennicutt et al.(2011)]{Ken11}Kennicutt, R.~C.\ 2011, \pasp,
  accepted (arXiv:1111.4438)

\bibitem[Koekemoer et al.(2002)]{Koe02}A. Koekemoer, A. Fruchter,
  R. Hook, W. Hack, 2002 HST Calibration Workshop, p. 337


\bibitem[Laor \& Draine(1993)]{Lao93} Laor, A.~\& Draine, B.~T.\ 1993,
  \apj, 402, 441  

\bibitem[Leibundgut et al.(1991)]{Lei91} Leibundgut, B., Kirshner,
  R.~P., Pinto, P.~A., et al.\ 1991, \apj, 372, 531

\bibitem[Leibundgut et al.(1993)]{Lei93} Leibundgut, B., Kirshner,
  R.~P., \& Porter, A.~C.\ 1993, Bulletin of the American Astronomical
  Society, 25, 834


\bibitem[Lucy et al.(1989)]{Luc89} Lucy, L.~B., Danziger, I.~J.,
  Gouiffes, C., \& Bouchet, P.\ 1989, IAU Colloq.~120: Structure and
  Dynamics of the Interstellar Medium, 350, 164


\bibitem[Makovoz \& Khan(2005)]{MK05} Makovoz, D., \& Khan, I.\ 2005,
  Astronomical Data Analysis Software and Systems XIV, 347, 81

\bibitem[Mathis et al.(1977)]{MRN77} Mathis, J.~S., Rumpl, W., \&
  Nordsieck, K.~H.\ 1977, \apj, 217, 425

\bibitem[Matsuura et al.(2011)]{Mat11} Matsuura, M., Dwek, E.,
  Meixner, M., et al.\ 2011, Science, 333, 1258

\bibitem[Michael et al.(2000)]{Mic00} Michael, E., McCray, R., Pun,
  C.~S.~J., et al.\ 2000, \apjl, 542, L53

\bibitem[Monet et al.(2003)]{Mon03} Monet, D.~G., Levine, S.~E.,
  Canzian, B., et al.\ 2003, \aj, 125, 984

\bibitem[Montes et al.(1998)]{Mon98} Montes, M.~J., van Dyk, S.~D.,
  Weiler, K.~W., Sramek, R.~A., \& Panagia, N.\ 1998, \apj, 506, 874

\bibitem[Morse et al.(2004)]{2004ApJ...614..727M} Morse, J.~A., Fesen, 
R.~A., Chevalier, R.~A., et al.\ 2004, \apj, 614, 727 

\bibitem[Nakar \& Sari(2010)]{Nak10}  Nakar, E.,  \& Sari,  R.\ 2010,
  \apj, 725, 904


\bibitem[Patat(2005)]{Pat05} Patat, F.\ 2005, \mnras, 357, 1161


\bibitem[Pozzo et al.(2004)]{Poz04} Pozzo, M., Meikle, W.~P.~S.,
  Fassia, A., et al.\ 2004, \mnras, 352, 457

\bibitem[Press et al.(1992)]{Pre92} Press, W.~H., Teukolsky, S.~A.,
  Vetterling, W.~T., \& Flannery, B.~P.\ 1992, Cambridge: University
  Press, 2nd ed.

\bibitem[Pun et al.(2002)]{Pun02} Pun, C.~S.~J., Michael, E., Zhekov,
  S.~A., et al.\ 2002, \apj, 572, 906

\bibitem[Rabinak \& Waxman(2011)]{Rab11} Rabinak, I., \& Waxman,
  E.\ 2011, \apj, 728, 63

\bibitem[Rajagopal et al.(2007)]{Raj07} Rajagopal, J., Menut, J.-L.,
  Wallace, D., et al.\ 2007, \apj, 671, 2017

\bibitem[Rowan-Robinson(1980)]{RR80} Rowan-Robinson, M.\ 1980, \apjs,
  44, 403

\bibitem[Rybicky \& Lightman(1985)]{RL85} Rybicky, G. B., \& Lightman,
  A. P.\ 1985, Radiative Processes in Astrophysics (New York: Wiley) 

\bibitem[Sahu et al.(2006)]{Sah06} Sahu, D.~K., Anupama, 
G.~C., Srividya, S., \& Muneer, S.\ 2006, \mnras, 372, 1315 


\bibitem[Sgro(1975)]{Sgro75} Sgro, A.~G.\ 1975, \apj, 197, 621

\bibitem[Sugerman(2003)]{Sug03} Sugerman, B.~E.~K.\ 2003, \aj, 126,
  1939

\bibitem[Sugerman(2005)]{S05} Sugerman, B.~E.~K.\ 2005, \apjl, 632,
  L17

\bibitem[Sugerman \& Crotts(2002)]{SC02} Sugerman, B.~E.~K., \&
  Crotts, A.~P.~S.\ 2002, \apjl, 581, L97

\bibitem[Sugerman et al.(2005)]{Sug05} Sugerman, B.~E.~K., Crotts,
  A.~P.~S., Kunkel, W.~E., Heathcote, S.~R., \& Lawrence, S.~S.\ 2005,
  \apjs, 159, 60

\bibitem[Sugerman et al.(2002)]{Sug02} Sugerman, B.~E.~K., Lawrence,
  S.~S., Crotts, A.~P.~S., Bouchet, P., \& Heathcote, S.~R.\ 2002,
  \apj, 572, 209




\bibitem[Uomoto \& Kirshner(1986)]{Uom86} Uomoto, A., \& Kirshner,
  R.~P.\ 1986, \apj, 308, 685

\bibitem[Uomoto(1991)]{Uom91} Uomoto, A.\ 1991, \aj, 101, 1275

\bibitem[V{\'a}rosi \& Dwek(1999)]{VD99} V{\'a}rosi, F., \& Dwek,
  E.\ 1999, \apj, 523, 265 (VD99)

\bibitem[Voors et al.(2000)]{Voo00} Voors, R.~H.~M., Waters,
  L.~B.~F.~M., de Koter, A., et al.\ 2000, \aap, 356, 501

\bibitem[Wang \& Chevalier(2002)]{WC02} Wang, C.-Y., \& Chevalier,
  R.~A.\ 2002, \apj, 574, 155

\bibitem[Warren et al.(2003)]{War03} Warren, J.~S., Hughes, J.~P., \&
  Slane, P.~O.\ 2003, \apj, 583, 260

\bibitem[Weiler et al.(1992)]{Wei92} Weiler, K.~W., van Dyk, S.~D.,
  Panagia, N., \& Sramek, R.~A.\ 1992, \apj, 398, 248

\bibitem[Weingartner \& Draine(2001)]{WD01}
  Weingartner, J.~C.~\& Draine, B.~T.\ 2001, \apj, 548, 296 (WD01)

\bibitem[Weis(2011)]{Wei11} Weis, K.\ 2011, IAU Symposium, 272, 372

\bibitem[Welch et al.(2007)]{Wel07} Welch, D.~L., Clayton, G.~C.,
  Campbell, A., et al.\ 2007, \apj, 669, 525

\bibitem[Wild \& Barbon(1980)]{Wil80} Wild, P., \& Barbon, R.\ 1980,
  \iaucirc, 3532, 1

\bibitem[Xu et al.(1995)Xu, Crotts \& Kunkel]{Xu95}Xu, J., Crotts,
  A.~P.~S.~\& Kunkel, W.~E.~1995, \apj, 451, 806 (Erratum: 463, 391)


\end{thebibliography}
\end{document}